\documentclass[]{IEEEtran}

\usepackage{cite}

\usepackage{graphicx}

\usepackage[cmex10]{amsmath}
\usepackage[utf8]{inputenc}
\usepackage[english]{babel}
\usepackage{amssymb}
\usepackage{amsthm}
\usepackage{mathtools}

\usepackage{algorithm}

\allowdisplaybreaks
\usepackage{algorithmic}

\usepackage{array}

\begin{document}

	\title{Granger Causality Detection via \\Sequential Hypothesis Testing}

\author{\IEEEauthorblockN{ Rahul Devendra, Ribhu Chopra, Kumar Appaiah}

\thanks{R. Devendra is with Microsoft Research, Bangalore, India. R. Chopra is with the Indian Institute of Technology Guwahati, Guwahati, Assam, India. K. Appaiah is with the Indian Institute of Technology Bombay, Mumbai, Maharashtra, India.
Emails: rahu760@gmail.com, ribhu@outlook.com, akumar@ee.iitb.ac.in
 }
}

	\maketitle

	\begin{abstract}
			Most of the metrics used for detecting a causal relationship among multiple time series ignore the effects of practical measurement impairments, such as finite sample effects, undersampling and measurement noise. It has been shown that these effects significantly impair the performance of the underlying causality test. In this paper, we consider the problem of sequentially detecting the causal relationship between two time series while accounting for these measurement impairments. In this context, we first formulate the problem of Granger causality detection as a binary hypothesis test using the norm of the estimates of the vector auto-regressive~(VAR) coefficients of the two time series as the test statistic. Following this, we investigate sequential estimation of these coefficients and formulate a sequential test for detecting the causal relationship between two time series. Finally via detailed simulations, we validate our derived results, and evaluate the performance of the proposed causality detectors.
		\end{abstract}

	\begin{IEEEkeywords}
		Causality Detection, Granger Causality, Sequential Detection, Hypothesis testing
	\end{IEEEkeywords}

		\section{Introduction}
The problem of detecting causal relationships between two or more time series is an important research problem across various scientific disciplines~\cite{cause1,Granger}, including, but not limited to neuroscience~\cite{Goebel2003,Hesse2003,Chen2004,Dhamala2008,Hu2012,Sampath2014,Hu2014,Seth3293}, physics~\cite{Ganapathy2007,Stramaglia2014}, climate science~\cite{Kodra2011}, econometrics~\cite{White2014}, etc. It has been shown that if the time series of interest are Gaussian, then the underlying causative relationship can be inferred using the Granger Causality Index~(GCI)~\cite{Granger,Price}. The GCI is the logarithm of ratio of the prediction error variances of caused time series with and without factoring in the effects of the causing time series. The GCI has been used extensively for causality detection across multiple research areas, including those listed above. Most of the present works dealing with causative relationships among time series assume exact knowledge of the second order statistics during the detection process. However, in making this assumption, these works ignore the effects of sampling noise and the finite sample effects due to the limited number of samples being used for estimating these statistics. Consequently, rather than the deterministic second order statistics, as assumed previously, the causality detector has to work with finite sample estimates of these quantities.

The authors in~\cite{CRM_TSP_2018} have addressed this problem by treating the available estimates of the second order statistics of the time series as random variables, and using these to derive the performance of the GCI in the presence of noise and finite sample effects. These performances have been derived in terms of the probabilities of detection and false alarm. The authors also derive two alternative analytically tractable test statistics in~\cite{CRM_TSP_2018} and show that these result in a performance comparable to the GCI. However, these test statistics, while incorporating the finite sample effects, require the use of all the samples for calculating them. Hence, these test statistics are unsuitable in a real time setting where the samples from the time series may arrive sequentially. Therefore, in this paper we propose a novel sequential test for detecting causal relationships among time series. Our key contributions in this direction are listed as follows.
\begin{enumerate}
	\item We propose to use the estimated vector autoregressive~(VAR) coefficients of the two time series as a test statistic for detecting the presence of a causative relationship between the two time series. We then analyze the performance of this test statistic in terms of the probabilities of detection and false alarm as a function of the number of samples of the time series.
	\item We then use the recursive least squares~(RLS) algorithm to develop a sequential test for causality detection among the two time series. 
	\item Via detailed simulations on artificially generated and real life data, we analyze the performance of the proposed VAR coefficient based test statistic. Following this, we evaluate the performance of the proposed sequential test and prescribe parameters for its optimal performance.
\end{enumerate}
The results developed in this paper can potentially be used to develop a real time test for detecting causal relationships among noisy time series. We next describe the system model considered in this work.
\section{System Model}
We consider the noisy measurements of two discrete time zero mean Gaussian random processes $u[n]$ and $v[n]$ having variances $\sigma_u^2$ and $\sigma_v^2$, respectively. We can jointly express $u[n]$ and $v[n]$ in the form of a bi-variate AR-$K$ model as
\begin{eqnarray}
	u[n]=\sum_{k=1}^{K} a_{uu,k}^*u[n-k]+\sum_{k=1}^{K}a^*_{uv,k}v[n-k]+\eta_u[n] \nonumber \\
	v[n]=\sum_{k=1}^{K} a^*_{vu,k}u[n-k]+\sum_{k=1}^{K}a^*_{vv,k}v[n-k]+\eta_v[n],
	\label{eq:bivariate_mod_scaler}
\end{eqnarray}
with  $a_{ij,k}$, $(i,j)\in\{u,v\}$ being the regression coefficients, $(\cdot)^*$ representing complex conjugation, and $\eta_u[n]$ and $\eta_v[n]$ being the innovation components corresponding to $u[n]$ and $v[n]$, respectively. We assume that $u[n]$ and $v[n]$ are sampled at the Nyquist rate, and the temporally white innovation process vector $\boldsymbol{\eta}[n]=[\eta_{u}[n]\, , \eta_{v}[n]]^T$ is distributed as $$\boldsymbol{\eta}[n]\sim\mathcal{CN}\left(0, \left[ \begin{array}{lr} \sigma_{\eta_{v}}^2&0 \\ 0 &\sigma_{\eta_{v}}^2 \end{array}\right] \right).$$ For simplicity, we assume unidirectional coupling, i.e. $a_{vu,k}=0\, ,\text{for} \, 1 \le k\le K$.
Now letting $\mathbf{u}_{K}[n]\triangleq[u[n],u[n-1],\dots,u[n-K+1]]^{T}$, $\mathbf{v}_{K}[n]\triangleq[v[n],v[n-1],\dots,v[n-K+1]]^{T}$, $\mathbf{a}_{ij}\triangleq[a_{ij,1},a_{ij,2},\dots,a_{ij,K}]^{T}\,, (i,j)\in\{u,v\}$, we can write~\eqref{eq:bivariate_mod_scaler} as,
\begin{equation}
	\left [ \begin{array}{c} u[n] \\ v[n] \end{array} \right] =\left [ \begin{array}{lr} \mathbf{a}_{uu}^H & \mathbf{a}_{uv}^H \\ \mathbf{0}^{H}_K & \mathbf{a}_{vv}^H \end{array} \right]\left [ \begin{array}{c} \mathbf{u}_K[n-1] \\ \mathbf{v}_K[n-1] \end{array} \right] + \left [ \begin{array}{c} \eta_{u}[n] \\ \eta_v[n] \end{array} \right] ,
	\label{eq:bivariate}
\end{equation}
with $\mathbf{0}_K$ being the $K$ dimensional all zero vector. Defining $r_{uu}[\tau]\triangleq E[u[n]u^{*}[n-\tau]]$, $r_{uv}[\tau]\ \triangleq E[u[n]v^{*}[n-\tau]]$, $r_{vu}[\tau]\triangleq E[v[n]u^{*}[n-\tau]]$ and $r_{vv}[\tau]\triangleq E[v[n]v^{*}[n-\tau]]$, and letting $\mathbf{r}_{uv,K}[\tau]=[r_{uv}[\tau],\ldots,r_{uv}[\tau-K+1]]^T$, etc., we can write,
\begin{equation}
	r_{uu}[\tau]=\mathbf{a}^{H}_{uu}\mathbf{r}_{uu,K}[\tau-1]+\mathbf{a}_{uv}^{H}\mathbf{r}_{vu,K}[\tau-1]+\sigma_{\eta_u}^2\delta[\tau].
	\label{eq:rpp}
\end{equation}
We can similarly write,
\begin{equation}
	r_{uv}[\tau]=\mathbf{a}^{H}_{uu}\mathbf{r}_{uv,K}[\tau-1]+\mathbf{a}_{uv}^{H}\mathbf{r}_{vv,K}[\tau-1]
	\label{eq:rpq}
\end{equation}
and
\begin{equation}
	r_{vv}[\tau]=\mathbf{a}_{vv}^{H}\mathbf{r}_{vv,K}[\tau-1]+\sigma_{\eta_v}^2\delta[\tau].
	\label{eq:rqq}
\end{equation}
The above equations, along with the the information that $r_{uu}[0]=\sigma_u^2$ and $r_{vv}[0]=\sigma_v^2$, can be used to compute $r_{uu}[\tau]$, $r_{uv}[\tau]$, and $r_{vv}[\tau]$ for different values of $\tau$.

Defining,
$
	\mathbf{R}_{uu,K}[\tau]\triangleq E\left[\mathbf{u}_{K}[n]\mathbf{u}_{K}^H[n-\tau]\right],
	\mathbf{R}_{uv,K}[\tau]\triangleq E\left[\mathbf{u}_{K}[n]\mathbf{v}_{K}^H[n-\tau]\right],
$ and similarly, $\mathbf{R}_{vu,K}[\tau]$ and $\mathbf{R}_{vv,K}[\tau]$,
it can be shown that~\cite{haykin_AFT},
\begin{equation}
	\left[\begin{array}{c}\mathbf{a}_{uu} \\ \mathbf{a}_{uv}\end{array}\right] = \left[\begin{array}{lr} \mathbf{R}_{uu,K}[0] & \mathbf{R}_{uv,K}[0] \\ \mathbf{R}_{vu,K}[0] & \mathbf{R}_{vv,K}[0]  \end{array}\right]^{-1}\left[\begin{array}{c}\mathbf{r}_{uu,K}[1] \\ \mathbf{r}_{uv,K}[1]\end{array}\right].
\end{equation}

It is evident from~\eqref{eq:bivariate} that the causal dependence of $u[n]$ over $v[n]$ depends on the coefficient vector $\mathbf{a}_{uv}$; that is, there exists a causal relationship between $u[n]$ and $v[n]$ only if $\mathbf{a}_{uv}\neq\mathbf{0}_K$. Therefore, similar to the GCI, the $\ell_2$ norm of $\mathbf{a}_{uv}$ can be used as a test statistic for the presence of a causal relationship between $u[n]$ and $v[n]$, with $\Vert \mathbf{a}_{uv}\Vert=0$ indicating the absence of a causal relationship and $\Vert \mathbf{a}_{uv}\Vert\neq0$ indicating its presence.
\section{Effects of Sampling Impairments}
Having described the test statistic previously, in this section we will investigate the effects of sampling impairments, viz. additive noise and finite sample effects, and use those to derive the underlying probabilities of detection and false alarm.
\subsection{Additive Noise}
Incorporating the effects of additive noise, we can define $x[n]$ and $y[n]$ as the noise corrupted versions of $u[n]$ and $v[n]$ respectively, such that,
\begin{equation}
	x[n]=u[n]+\nu_x[n], \quad y[n]=v[n]+\nu_y[n]
	\label{eq:addnoise}
\end{equation}
with $\nu_i[n]\sim\mathcal{CN}(0,\sigma_{\nu_i}^2)$, $i\in\{x,y\}$. Hence, if a causal relationship exists between, $u[n]$ and $v[n]$, then a causal relationship also exists between $x[n]$ and $y[n]$. We can therefore express $x[n]$ as,
 \begin{equation}
 	x[n]  = [ \mathbf{w}_{x}^H  \mathbf{w}_{y}^H ]\left[ \begin{array}{c} \mathbf{x}_K[n-1] \\ \mathbf{y}_K[n-1] \end{array} \right] +  \varphi[n],
 \end{equation}
with $\varphi[n]$ representing the zero mean prediction error for $x[n]$ in terms of $\mathbf{x}_K[n-1]$ and $\mathbf{y}_K[n-1] $, and $\mathbf{w}_{x}^H$ and $\mathbf{w}_{y}^H$ regression weights that minimize the mean squared value of $\varphi[n]$. It is easy to show that~\cite{haykin_AFT},
\begin{multline}
	\left[\begin{array}{c}\mathbf{w}_{x} \\ \mathbf{w}_{y}\end{array}\right] = \left[\begin{array}{lr} \mathbf{R}_{xx,K}[0] & \mathbf{R}_{xy,K}[0] \\ \mathbf{R}_{yx,K}[0] & \mathbf{R}_{yy,K}[0]  \end{array}\right]^{-1}\left[\begin{array}{c}\mathbf{r}_{xx,K}[1] \\ \mathbf{r}_{xy,K}[1]\end{array}\right]\\
	=\left[\begin{array}{lr} \mathbf{R}_{uu,K}[0]+\sigma_{\nu_x}^2\mathbf{I}_K & \mathbf{R}_{uv,K}[0] \\ \mathbf{R}_{vu,K}[0] & \mathbf{R}_{vv,K}[0]+\sigma_{\nu_y}^2\mathbf{I}_K  \end{array}\right]^{-1}\left[\begin{array}{c}\mathbf{r}_{uu,K}[1] \\ \mathbf{r}_{uv,K}[1]\end{array}\right].
\end{multline}
We can also use the results in~\cite{haykin_AFT} to show that the minimum mean squared value of $\varphi[n]$ is given by
$	\sigma_{\varphi}^2=\sigma_{\eta_v}^2+\sigma_{\nu_y}^2.$
Now, if a causal relationship exists between $x[n]$ and $y[n]$, then $\mathbf{w}_y$ will be a nonzero vector, otherwise it will be an all zero vector. Hence, the norm of  $\mathbf{w}_y$ can still be used as a test statistic for determining the existence of a causal relationship between two noisy time series. We next determine the effects of using estimated second order statistics on this test statistic.
\subsection{Finite Sample Effects}
We now consider the case where $N$ samples each of $x[n]$ and $y[n]$ are observed, and are used to calculate the test statistics. We note that in the absence of the availability of exact second order statistics, we cannot use the minimum mean squared error~(MMSE) estimator for $\mathbf{w}_x$ and $\mathbf{w}_y$, and will have to use the least squares~(LS) estimator~\cite{haykin_AFT}. Defining the data matrix $\mathbf{A}[N]$ as,
\begin{equation}
\mathbf{A}[N]=\left[\begin{array}{ccc}\mathbf{x}_{K}[N]&\ldots&\mathbf{x}_{K}[K+1] \\ \mathbf{y}_{K}[N]&\ldots&\mathbf{y}_{K}[K+1]\end{array}\right]^H=\left[\begin{array}{c}\mathbf{A}_{x}[N] \\ \mathbf{A}_{y}[N]\end{array}\right],
\end{equation}
we can write the estimate of the auto-correlation matrix as
\begin{align} \boldsymbol{\Phi}[N]&=\frac{1}{N-K}\mathbf{A}^H[N]\mathbf{A}[N] \nonumber
	\\&=\frac{1}{N-K}\left[\begin{array}{cc}\mathbf{A}_{x}^H[N]\mathbf{A}_x[N]& \mathbf{A}_{x}^H[N]\mathbf{A}_y[N]\\ \mathbf{A}_{y}^H[N]\mathbf{A}_x[N] & \mathbf{A}_{y}^H[N]\mathbf{A}_y[N]\end{array}\right] \nonumber
		\\&= \left[\begin{array}{cc}\boldsymbol{\Phi}_{xx}[N]& \boldsymbol{\Phi}_{xy}[N]\\ \boldsymbol{\Phi}_{yx}[N] & \boldsymbol{\Phi}_{yy}[N]\end{array}\right]
	\label{eq:block_mat}
\end{align} and the estimate of the cross-correlation vector as
\begin{multline}
\boldsymbol{\psi}[N]=\frac{1}{N-K}\mathbf{A}^H[N]\mathbf{x}_{N-K}[N]\\=\frac{1}{N-K}\left[\mathbf{A}^H_x[N]\mathbf{A}^H_y[N]\right]\mathbf{x}^H_{N-K}[N]=\left[\begin{array}{c}\boldsymbol{\psi}_x[N]\\\boldsymbol{\psi}_y[N]\end{array}\right].
\end{multline}
Consequently, we can write the $N$ sample LS estimate of the weight vector as
\begin{multline}
	\left[\begin{array}{c}\hat{\mathbf{w}}_{x}[N] \\ \hat{\mathbf{w}}_{y}[N]\end{array}\right]=\boldsymbol{\Phi}^{-1}[N]\boldsymbol{\psi}[N]\\=\left[\begin{array}{cc}\boldsymbol{\Phi}_{xx}[N]& \boldsymbol{\Phi}_{xy}[N]\\ \boldsymbol{\Phi}_{yx}[N] & \boldsymbol{\Phi}_{yy}[N]\end{array}\right]^{-1}\left[\begin{array}{c}\boldsymbol{\psi}_x[N]\\\boldsymbol{\psi}_y[N]\end{array}\right].
\end{multline}
Now, we can express $\hat{\mathbf{w}}_y$ as,
\begin{equation}
	\hat{\mathbf{w}}_y=\boldsymbol{\Gamma}_1[N]\boldsymbol{\psi}_x[N]+\boldsymbol{\Gamma}_2[N]\boldsymbol{\psi}_y[N],
\end{equation}
where, $\boldsymbol{\Gamma}_1[N]=\left(\boldsymbol{\Phi}_{xy}[N]-\boldsymbol{\Phi}_{xx}[N]\boldsymbol{\Phi}_{yx}^{-1}[N]\boldsymbol{\Phi}_{yy}[N]\right)^{-1}$ and $\boldsymbol{\Gamma}_2[N]=\left(\boldsymbol{\Phi}_{yy}[N]-\boldsymbol{\Phi}_{yx}[N]\boldsymbol{\Phi}_{yy}^{-1}[N]\boldsymbol{\Phi}_{xy}[N]\right)^{-1}$.

Now, since $\hat{\mathbf{w}}_y$, represents the LS estimate of $\mathbf{w}_y$ and can be expressed as~\cite{haykin_AFT},
\begin{equation}
	\hat{\mathbf{w}}_y={\mathbf{w}}_y+\tilde{\mathbf{w}}_y,
\end{equation}
where $\tilde{\mathbf{w}}_y$ is the estimation error vector that can be shown to be distributed as $\tilde{\mathbf{w}}_y\sim\mathcal{CN}\left(\mathbf{0}_K,\frac{\sigma_{\varphi}^2}{N-K}\boldsymbol{\Sigma}\right)$, with $\mathbf{0}_K$ being the all zero vector of length $K$ and \begin{equation}
	\boldsymbol{\Sigma}=\left[\boldsymbol{\bar{\Gamma}}_1\;\boldsymbol{\bar{\Gamma}}_2\right]\left[\begin{array}{lr} \mathbf{R}_{xx,K}[0] & \mathbf{R}_{xy,K}[0] \\ \mathbf{R}_{yx,K}[0] & \mathbf{R}_{yy,K}[0]  \end{array}\right] \left[\begin{array}{c} \boldsymbol{\bar{\Gamma}}^H_1  \\ \boldsymbol{\bar{\Gamma}}^H_2 \end{array}\right],
\end{equation}
with $\boldsymbol{\bar{\Gamma}}_1=\left(\mathbf{R}_{xy,N}[0]-\mathbf{R}_{xx,N}[0]\mathbf{R}_{yx,N}^{-1}[0]\mathbf{R}_{yy,N}[0]\right)^{-1}$ and $\boldsymbol{\bar{\Gamma}}_2=\left(\mathbf{R}_{yy,N}[0]-\mathbf{R}_{yx,N}[0]\mathbf{R}_{yy,N}^{-1}[0]\mathbf{R}_{xy,N}[0]\right)^{-1}$.

Since the entries of $\tilde{\mathbf{w}}_y$ and hence $\hat{\mathbf{w}}_y$ are not independent and identically distributed~(i.i.d.), it is not possible to determine the statistics of $\Vert\hat{\mathbf{w}}_y\Vert$ in a closed form. This results in the analysis of $\Vert\hat{\mathbf{w}}_y\Vert$ as a test statistic for the causal relationship between $x[n]$ and $y[n]$ becoming intractable.

Defining $\mathbf{v}=\sqrt{\frac{N-K}{\sigma_{\varphi}^2}}\boldsymbol{\Sigma}^{-\frac{1}{2}}\mathbf{\hat{w}}_y$, we obtain
\begin{equation}
\mathbf{v}=	\frac{1}{\sigma_{\varphi}}\sqrt{N-K}\boldsymbol{\Sigma}^{-\frac{1}{2}}{\mathbf{w}}_y+\tilde{\mathbf{v}},
\end{equation}
with $\tilde{\mathbf{v}}\sim\mathcal{CN}\left(\mathbf{0}_K,\mathbf{I}_K\right)$, and hence ${\mathbf{v}}\sim\mathcal{CN}\left(\sqrt{\frac{N-K}{\sigma_{\varphi}^2}}\boldsymbol{\Sigma}^{-\frac{1}{2}}{\mathbf{w}}_y,\mathbf{I}_K\right)$. Since the entries of $\mathbf{v}$ are i.i.d. Gaussian with unit variance, $\Vert\mathbf{v}\Vert^2$ becomes a Chi-squared distributed random variable with $2K$ degrees of freedom and a non centrality parameter $\kappa=\frac{N-K}{\sigma_{\varphi}^2}\mathbf{w}_y^H\boldsymbol{\Sigma}^{-1}\mathbf{w}_y^H$.

Now, if a causal relationship does not exist between $x[n]$ and $y[n]$, then $\kappa$ becomes zero, making $\Vert\mathbf{v}\Vert^2$ a central Chi-squared rv. Consequently, if we denote the absence of a causal relationship by the null hypothesis, $\mathcal{H}_0$, and its presence by the alternate hypothesis $\mathcal{H}_1$, then we can write the binary hypothesis test for detecting a causal relationship between $x[n]$ and $y[n]$ in terms of the test statistic $T_N=\Vert\mathbf{v}\Vert^2$ as
\begin{equation}
	T_N\sim\left\{ \begin{array}{lr} \chi^2_{2K}(0)&\mathcal{H}_0 \\ \chi^2_{2K}(\kappa)&\mathcal{H}_1 \end{array}\right..
\end{equation}

Therefore, for a detection threshold $\lambda$, the event corresponding to the correct detection of the causal relationship between $x[n]$ and $y[n]$ can be expressed as,
\begin{equation}
	P_d=\Pr\{T_N>\lambda|\mathcal{H}_1\}=Q_{K}(\sqrt{\kappa},\sqrt{\lambda}),
\end{equation}
with $Q_K(\cdot,\cdot)$ being the Marcum Q-function of order $K$. Similarly, the probability of detecting a causal relationship between $x[n]$ and $y[n]$ when there is none, becomes
\begin{equation}
	P_{fa}=\Pr\{T_N>\lambda|\mathcal{H}_0\}=Q_{K}(0,\sqrt{\lambda}).
\end{equation}

Since the Marcum $Q$-function is an increasing function of the first argument, an increase in the number of samples, that results in an increased value of $\kappa$ improves the detection performance, which is consistent with intuition. However, this test statistic still processes the samples from the two time series as a block, and is unsuitable for sequential detection of a causal relationship between $x[n]$ and $y[n]$. Therefore, in the next section, we first describe an adaptive technique for sequentially updating the test statistic, and then use that to develop a sequential test for cauasality between $x[n]$ and $y[n]$.
\section{The Sequential Test}
We note that the weight vector $\mathbf{\mathbf{\hat{w}}}$ can be iteratively updated as $\mathbf{\mathbf{\hat{w}}}[n]$ at the $n$th instant using the recursive least squares~(RLS) algorithm. For this, we first set $\mathbf{\hat{w}}[0]=\mathbf{0}_K$, and $\mathbf{P}[0]=\delta\mathbf{I}_{2K}$. Following this, at the $n$th instant, we can first calculate the prediction error in $x[n]$ as,
\begin{equation}
	\tilde{{x}}[n]={x}[n]-[ \mathbf{\hat{w}}_{x}^H[n-1]  \mathbf{\hat{w}}_{y}^H[n-1] ]\left[ \begin{array}{c} \mathbf{x}_K[n-1] \\ \mathbf{y}_K[n-1] \end{array} \right],
\end{equation}
 	followed by an update in the sample covariance matrix, as
 	\begin{equation}
 		\boldsymbol{\Phi}[n]=\mu\boldsymbol{\Phi}[n-1]+\left[ \begin{array}{c} \mathbf{x}_K[n-1] \\ \mathbf{y}_K[n-1] \end{array} \right]\left[  \mathbf{x}^H_K[n-1] \quad \mathbf{y}^H_K[n-1]  \right],
 	\end{equation}
 with $0<\mu\le1$ being the forgetting factor. We can similarly update the cross correlation vector as,
 \begin{equation}
 	\boldsymbol{\psi}[n]=\mu\boldsymbol{\psi}[n-1]+\left[ \begin{array}{c} \mathbf{x}_K[n-1] \\ \mathbf{y}_K[n-1] \end{array} \right]  \mathbf{x}^*_K[n],
 \end{equation} 
this is followed by the calculation of the RLS gain as
\begin{equation}
 \mathbf{g}[n]=\frac{\boldsymbol{\Phi}^{-1}[n-1]\left[ \begin{array}{c} \mathbf{x}_K[n-1] \\ \mathbf{y}_K[n-1] \end{array} \right]}{\mu+\left[  \mathbf{x}^H_K[n-1] \quad \mathbf{y}^H_K[n-1]  \right]\boldsymbol{\Phi}^{-1}[n-1]\left[ \begin{array}{c} \mathbf{x}_K[n-1] \\ \mathbf{y}_K[n-1] \end{array} \right]}.
\end{equation}
and the update of $\boldsymbol{\Phi}^{-1}[n]$ as,
 \begin{multline}
 	\boldsymbol{\Phi}^{-1}[n]=\mu^{-1}\boldsymbol{\Phi}^{-1}[n-1]\\-\mu^{-1}\mathbf{g}[n]\left[  \mathbf{x}^H_K[n-1] \quad \mathbf{y}^H_K[n-1]  \right]\boldsymbol{\Phi}^{-1}[n-1],
 \end{multline}
finally leading to an updated estimate of the weight vector as,
\begin{equation}
\mathbf{\hat{w}}[n]	=\mathbf{\hat{w}}[n-1]+\mathbf{g}[n]\tilde{x}[n].
\end{equation}
This updated weight vector can be used to calculate the test statistic at the $n$th instant as, $T[n]=\frac{n-K}{\sigma_{\varphi}^2}\Vert\boldsymbol{\Sigma}^{-\frac{1}{2}}\mathbf{\hat{w}}[n]\Vert^2$. We then declare $x[n]$ to be caused by $y[n]$ if $T[n]>\lambda_1[n]$ and declare $x[n]$ to be not caused by $y[n]$ if $T[n]<\lambda_0[n]$. We wait for the next sample if $T[n]\in(\lambda_0[n],\lambda_1[n])$.
\section{Numerical Results}
We now present numerical results to corroborate our derived results and illustrate the performance of the derived test statistic for detecting a causal relationship between time series. For illustrating the efficacy of the derived test statistics, we consider a simple VAR-1 process modeled as
  \begin{equation}
  	u[n]= av[n-1]+\eta_u[n],\;
  	v[n]=\eta_v[n]
  	\label{eq:toy_model}
  \end{equation}
  with $a=0.25$, and generate its samples for different additive noise variances according to~\eqref{eq:addnoise}. We then evaluate the probabilities of detection and false alarm by averaging over 10,000 independent realizations of the signals of interest.

	\begin{figure}[t]
	\centering
	\includegraphics[width=.45\textwidth]{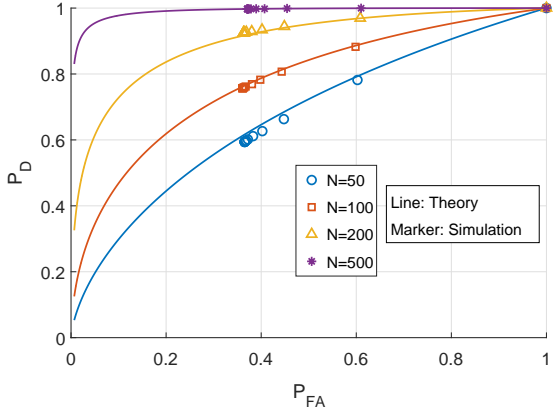}
	\caption{ROC for detection based on $T_N$, at an SNR of $0$~dB, for different numbers of samples.}
	\label{fig:ROC}
\end{figure}
In Fig.~\ref{fig:ROC}, we compare our derived results against the simulated receiver operating characteristics~(ROCs) for the example case discussed above for different numbers of samples at an SNR of $0$~dB. It is observed that for a small number of samples, both the simulated and theoretical the ROCs are close to the $P_{FA}=P_{D}$ line. However, the concavity of the ROC increases with an increase in the number of samples, as per intuition.

\begin{figure}[t]
	\centering
	\includegraphics[width=.45\textwidth]{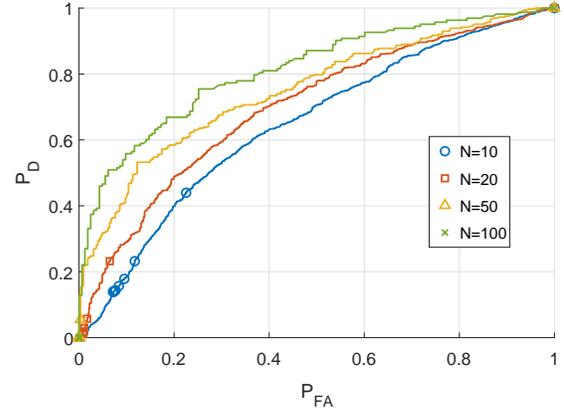}
	\caption{Performance of the test statistic $T_N$ on the real-world dataset $69$ in~\cite{JMLR:v17:14-518} for different numbers of samples for a data SNR of 20 dB.}
	\label{fig:ROC2}
\end{figure}

In Fig.~\ref{fig:ROC2} we plot the ROC for our proposed test statistic on a real-world dataset with known ground truth, made available in~\cite{JMLR:v17:14-518} for a data SNR of 20~dB for different window sizes~($N$). The figure corresponds to data pair $69$ of the database, containing temperature measurements from inside and outside a room taken every five minutes. This data set was selected out of the $108$ available data sets because it represented time series data that fit well into our proposed system model, and the number of samples in these time series was sufficiently large to obtain probabilities of detection and false alarm with a fair amount of accuracy.

\begin{figure}[t]
	\centering
	\includegraphics[width=.45\textwidth]{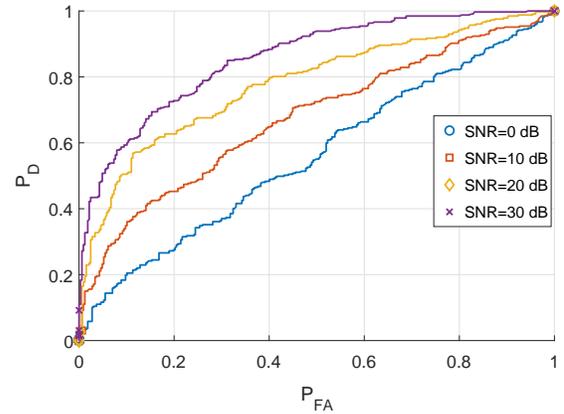}
	\caption{Performance of the test statistic $T_N$ on the real-world dataset $69$ in~\cite{JMLR:v17:14-518} for different data SNR at $N$=50.}
	\label{fig:ROC3}
\end{figure}

We note that neither the underlying model parameters of the data nor the variance of the additive noise are known. As a consequence, we fit this data into a VAR-1 model similar to the example considered previously, and assume it to be inherently noiseless. We first process these time series to make them zero mean. As a next step, we artificiality inject additive white Gaussian noise to obtain a given signal to noise ratio for the said time series. Following this, the data pairs are divided into windows of length $N$, and the test statistic is calculated and compared to the threshold. The causal relationship is marked as successfully detected if the test statistic exceeds the threshold, and the detection of the causal relationship is marked as missed otherwise. Simultaneously, we consider the noise injected into the time series separately and use it to generate the test statistic and compare it against the same threshold. In this case the test statistic exceeding the threshold implies a false alarm. The results of these comparisons are averaged over all the sample windows to obtain the probabilities of detection and false alarm. The jitters in the plots are due to the averaging over the relatively small number of samples that were available.  In Fig.~\ref{fig:ROC3} we repeat the same experiment as in Fig.~\ref{fig:ROC2} for a window size $N=50$ at different data SNRs.

\section{Conclusions}
In this work, we have derived, evaluated and demonstrated a novel test statistic for detecting causal relationships among time series. We have derived the underlying probabilities of detection and false alarm for the above mentioned test statistic and extended it to support sequential detection of causative relationships among time series. Via extensive numerical simulations we have shown that the derived results match well with simulated as well as real world data. Future work would consider the extension of these results to multiple undersampled time series.
 	\bibliographystyle{ieeetr}
\bibliography{lett_seq_cause}

\end{document}